\documentclass[12pt]{iopart}
\usepackage{graphicx}
\usepackage{float}

\begin{document}

\title[]{\bf Measurement of Non-photonic Electrons
in p + p Collisions at $\sqrt{s_{NN}}$ = 200 GeV with reduced
detector material in STAR}

\author{\bf F.Jin for the STAR Collaboration  $^{\rm a,\rm b}$}

\address{$^{\rm a}$Shanghai Institute of Applied Physics,
Chinese Academy of Sciences,P.O. Box 800-204, Shanghai 201800,
China }

\address{$^{\rm b}$Brookhaven National Laboratory, Upton, NY 11973, USA
}
\vspace{3pt}
\address{Email: \it jfazj@rcf.rhic.bnl.gov}

\begin{abstract}

In this paper, we present our analysis of mid-rapidity non-photonic
electron (NPE) production at $p_{T}$$>$0.2GeV/$c$ in p+p collisions at
$\sqrt{s_{NN}}$ = 200 GeV. The dataset is $\scriptsize{\sim}$78M
TOF-triggered events taken from RHIC year 2008 runs. Through the
measurement of $e/\pi$ ratio, we find that the photonic background
electrons from $\gamma$ conversions are reduced by about a factor of
10 compared with those in STAR previous runs due to the absence of
inner tracking detectors and the supporting materials. The dramatic
increase of signal-to-background ratio will allow us to improve the
precision on extracting the charm cross-section via its
semi-leptonic decays to electrons.

\end{abstract}

\submitto{\JPG}

\section{Introduction}\label{intro}

Ultra-relativistic heavy ion collisions can provide sufficient
conditions for the formation of a deconfined plasma of quarks and
gluons. Heavy-flavor quarks(charm and bottom) are produced
dominantly through high-$Q^{2}$ partonic interactions~\cite{STAR_0607012}.
Because of the large mass, it's expected that the cross-section
of heavy flavor production can be calculated
in perturbative quantum chromodynamics (pQCD)~\cite{MC}.
Precise measurements of charm total cross-section and
transverse momentum spectrum in p+p collisions
will provide a baseline to understand the charm production and in-medium
mechanism in heavy ion collisions~\cite{BW}.

To date, one way to study heavy flavor production
is to measure NPE production from their semi-leptonic decay.
Although the systematic uncertainties are quite large,
the charm cross-section measured by STAR is different from
that measured by PHENIX by a factor $\scriptsize{\sim}$2 or
1.5$\sigma$~\cite{STAR_0607012,Phenix_0802}. STAR has large and
uniform acceptance, but the material close to beam pipe in previous
run is
$\scriptsize{\sim}$5.5\% of a radiation length ($X_{0}$). In run8,
STAR removed inner tracking detectors, SVT (Silicon Vertex Tracker)
and SSD (Silicon Strip Detector). The material budget integrating
from interaction point to TPC inner field cage is
$\scriptsize{\sim}$0.69\%$X_{0}$. There are wraps around the beam
pipe to bake out the beam pipe and glues at inner field cage, which
are estimated. The exact material in terms of radiation length is
mapped from the data.

During the 2008 RHIC runs, TPC has upgraded the electronics of one
of its 24 sectors to a factor of 10 faster with negligible dead time
using a pipeline buffer (TPX in DAQ1000) ~\cite{tonko}. Fully
instrumentation of the 24 sectors will been completed after run8.
Five trays of Time-of-Flight (TOF) was placed behind the TPX sector,
and each tray covers -1$<$$|\eta|$$<$0 in pseudo-rapidity and
$\Delta\phi$$<$$\pi/$30 in azimuth. Two pVPDs were installed to
provide a starting time for TOF detectors, each staying 5.4m away
from the TPC center along the beam line. The starting time
resolution is $\scriptsize{\sim}$83ps. The timing resolution of TOF
is $\scriptsize{\sim}$110ps in p+p collisions. In our analysis, we
collected $\scriptsize{\sim}$78M TPX+TOF triggered events by
requiring at least one hit in TOF, equivalent to
$\scriptsize{\sim}$400M minimum bias events.

Due to STAR's unique capability of identifying electron and pion at
low $p_T$ in an identical procedure by a combination of ionization
energy loss $dE/dx$ in TPC and velocity $\beta$ from
TOF~\cite{mingshao}, many of the systematic uncertainties associated
with individual charge pion and electron cancel. In the analyses
presented here, we'll focus on the $e/\pi$ ratio and compare the
ratio from previous measurements and background. After selection of
good runs and a vertex cut of $|z_{vtx}|$$<$40cm to reject
conversion background electrons from beam pipe and its support, we
also need a rapidity cut -0.6$<$y$<$0 to ensure rapidity
distribution of electron similar as that of pion when we calculate
$e/\pi$ ratio. In order to get good primary track, we have a vertex
Z difference cut, $|z_{vtx}$(pVPD)-$z_{vtx}$(TPC)$|$$<$6cm.

\section{Particle (electron and pion) Identification}\label{eID}
\begin{figure}[htbp]
\resizebox{0.5\textwidth}{!}{\includegraphics{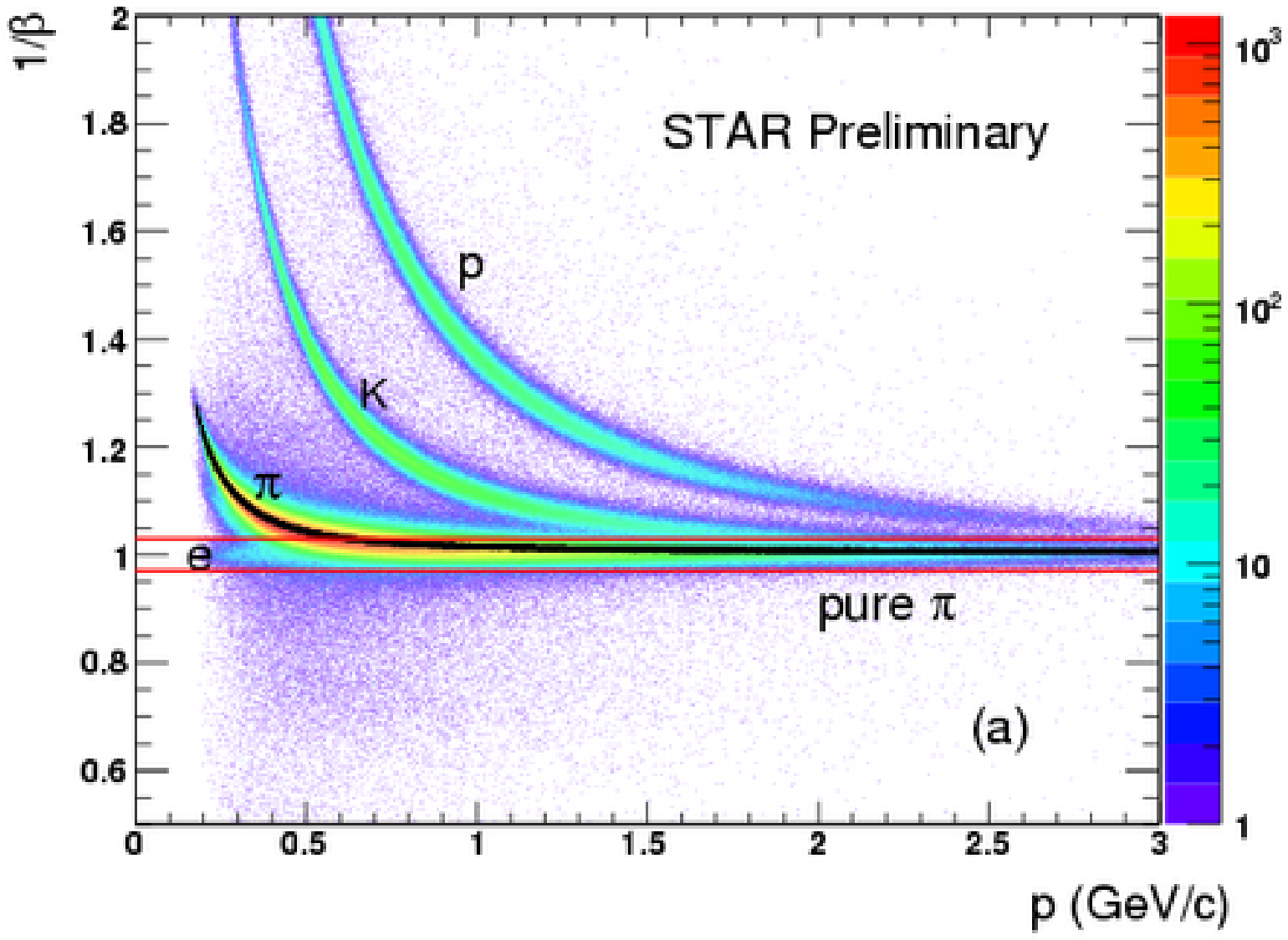}}
\resizebox{0.5\textwidth}{!}{\includegraphics{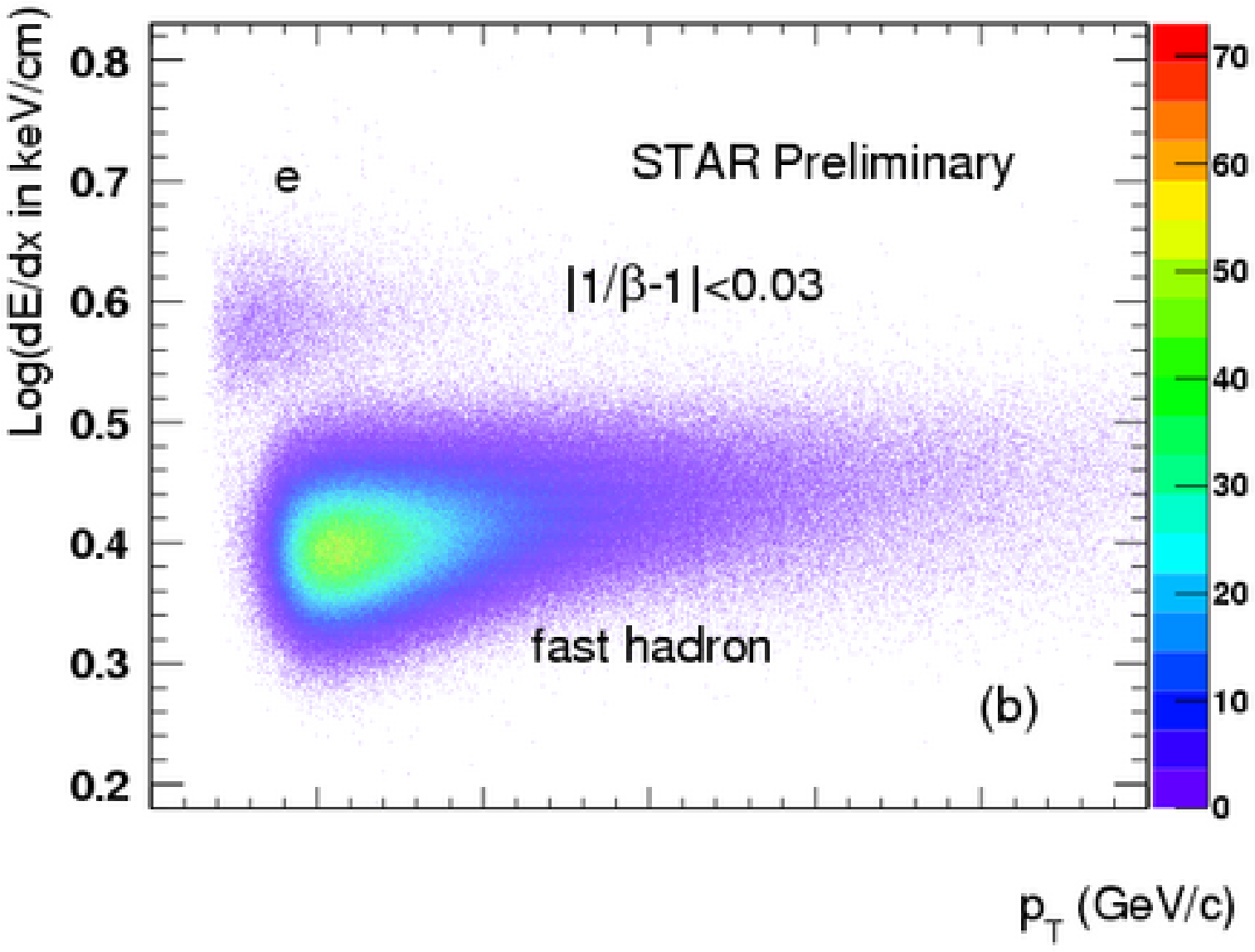}}
\caption[]{(Color online)(a) particle inverse velocity 1/$\beta$ as
a function of momentum $p$. The zone between two red solid lines
stands for TOF PID cut,$|$1/$\beta$-1$|$$<$0.03. The black belt is pure
$\pi$ sample to make sure the tail shape of fast hadrons. (b) $dE/dx$
of electron and fast hadrons versus transverse momentum $p_{T}$. }
\label{fig1}
\end{figure}

The relativistic rise of the $dE/dx$ in TPC from electrons
provides a possible separation of electrons from the rest of the
hadrons except $dE/dx$ from slow hadrons impinging the electron
$dE/dx$ band at several crossing points as function of momentum.
Electron identification requires TOF PID cut,
$|$1/$\beta$-1$|$$<$0.03, to reject slow hadrons. This cut is
shown in the Fig~\ref{fig1} (a) using two red solid lines. After
this cut we will get the $dE/dx$ distribution of electron and fast
hadrons as a function of $p_{T}$, shown in Fig~\ref{fig1} (b).
Projecting this plot in different $p_{T}$ bins and using suitable
function fit to $dE/dx$ distribution, we obtain the raw yields of
electron.

\begin{figure}[htbp]
\resizebox{0.5\textwidth}{!}{\includegraphics{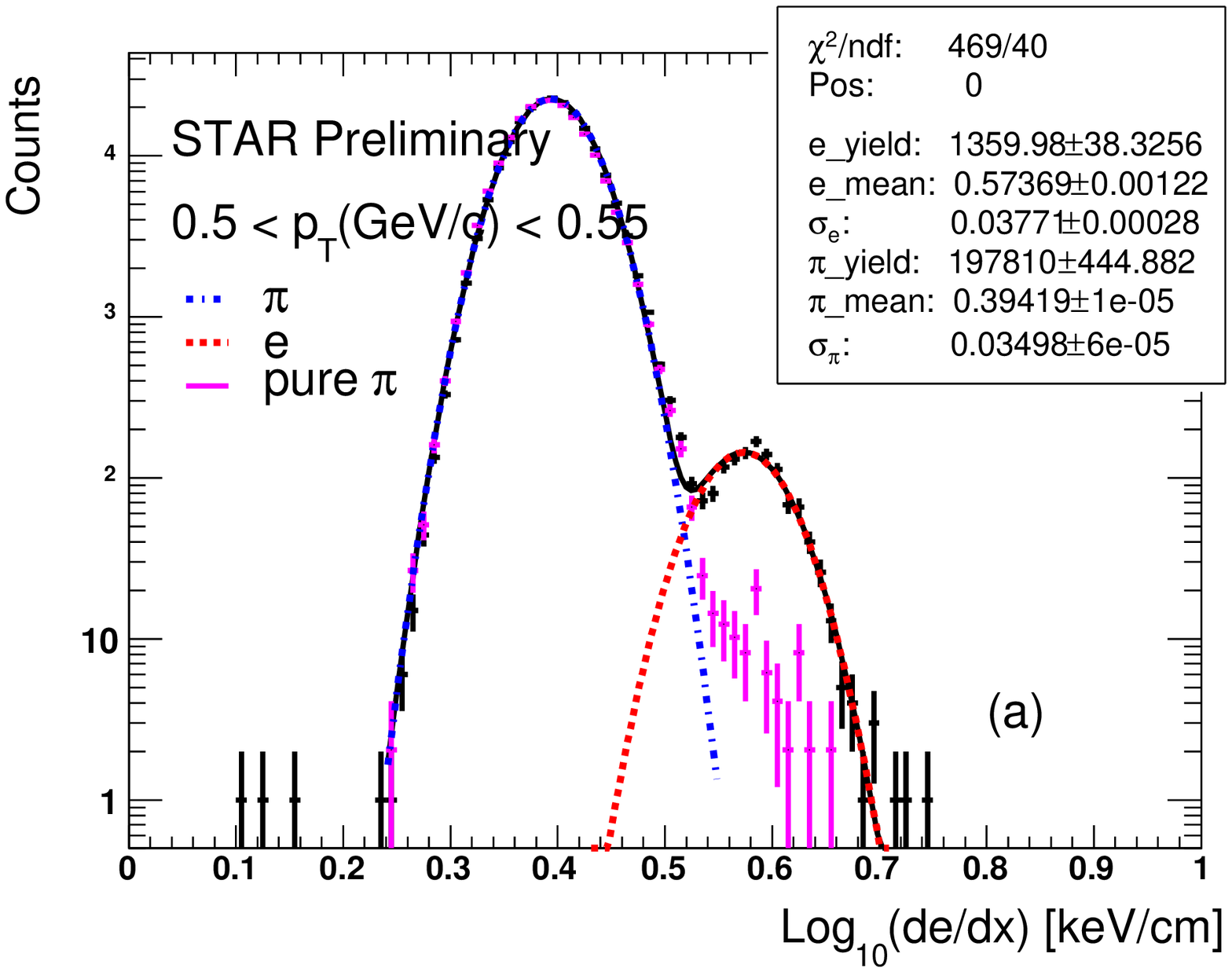}}
\resizebox{0.5\textwidth}{!}{\includegraphics{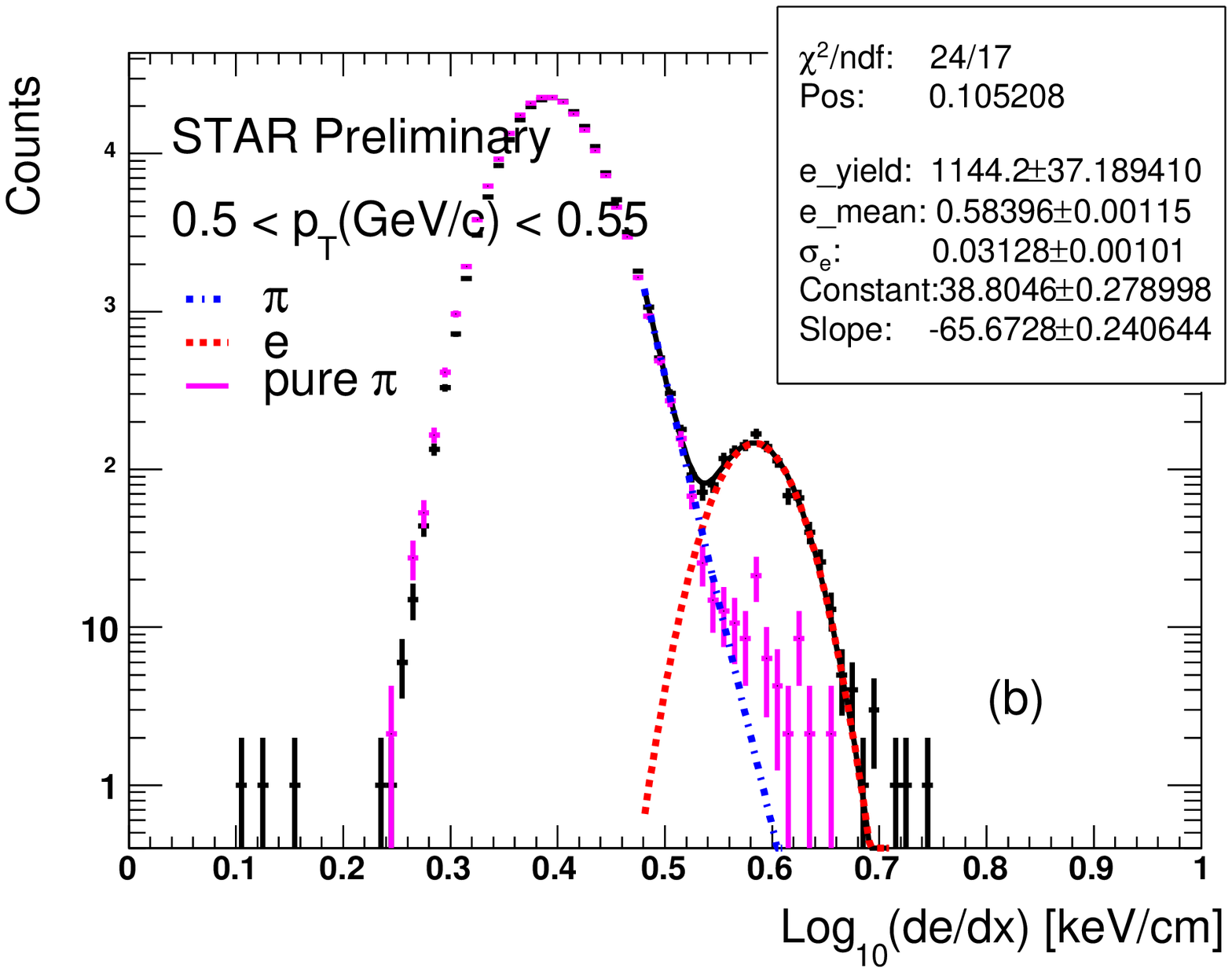}}
\caption[]{(Color online)Red dashed line stands for electron $dE/dx$
distribution; blue dotted-dashed line stands for fast hadrons (In our
analysis, it's $\pi$ below $p_{T}$$<$1.6GeV/$c$); pink histogram
stands for pure $\pi$.(a) the black solid line stands for 2-Gaussian
fit. It can't describe the overlap well. (b) the black solid line
stands for exponential+Gaussian fit. It fits the tail shape of fast
hadrons, the peak and width of electron and their overlap well. }
\label{fig2}
\end{figure}

We used two function forms to estimate the background $dE/dx$
distribution.
One is a Gaussian function and the other is an exponential function.
We found the 2-Gaussian function cannot describe the left shoulder
region of electron $dE/dx$ due to the tail of the $dE/dx$ from fast
hadron in lower $p_T$ region($p_{T}$$<$1.6GeV/$c$). Instead, a
function of exponential+Gaussian was used in the fit. We also use
two methods to produce the background tail shape of fast hadrons and
evaluate the uncertainty due to hadron contamination: 1)inverse
velocity difference between measurement and calculation
0$<$1/$\beta$(measured)-1/$\beta(\pi)$$<$0.01 to provide pure pion
$dE/dx$ distribution; 2)energy deposited in EMC $E$$<$0.5GeV to enhance
hadrons with non-electromagnetic showers. Figure~\ref{fig2} shows
the $dE/dx$ distribution together with a background distribution from
method 1) in 0.5$<$$p_{T}$$<$0.55(GeV/$c$). The exponential background
tail can reproduce the background distribution very well.

In 1.6$<$$p_{T}$$<$4.0(GeV/$c$), 3-Gaussian fit is used, assuming that
one Gaussian function can describe kaons and some of the residual
protons after the velocity cut. In order to check that we have a
real electron signal, we used energy deposited in EMC to reject
hadrons. Comparing the peak and width of electron with EMC and
without EMC selection, we can evaluate the electron yields at high
$p_{T}$.

\begin{figure}[htbp]
\resizebox{0.5\textwidth}{!}{\includegraphics{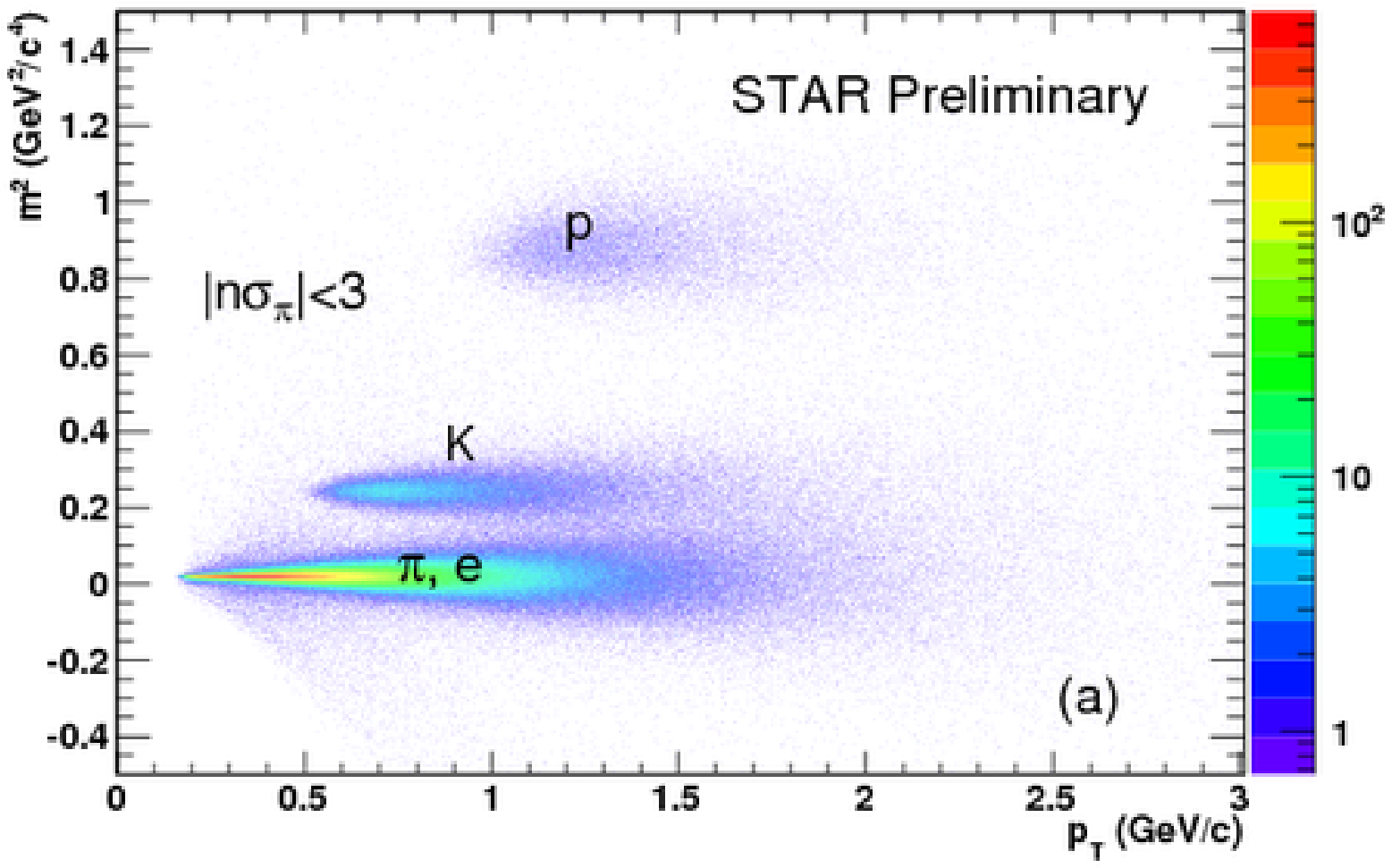}}
\resizebox{0.5\textwidth}{!}{\includegraphics{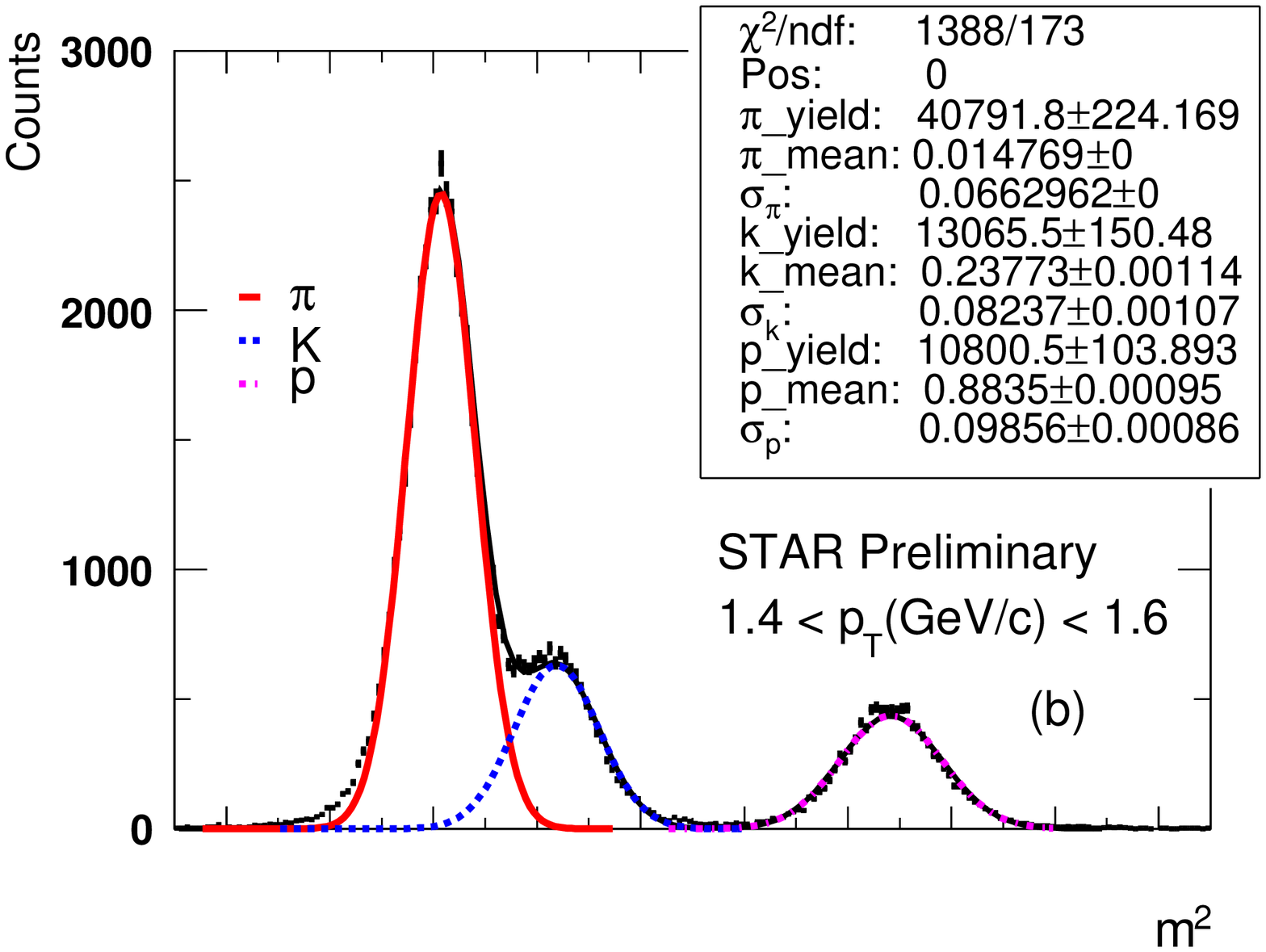}}
\caption[]{(Color online)(a) $m^2$ from TOF as a function of $p_T$.
$\pi$ is the dominant particle when $m^{2}$ is $\scriptsize{\sim}0$.
(b) Example of the $m^2$ distribution in a given $p_T$ bin.
Red solid line is fit for $\pi$; blue dashed line for K; pink
dotted-dashed line for p.
} \label{fig3}
\end{figure}

$\pi$ identification was achieved by a combination of $dE/dx$
$|n\sigma_{\pi}|$$<$3 and particle mass from TOF measurement via
$m^2 =p^2/(\beta\gamma)^2$. Fig~\ref{fig3} (a) shows particle mass
square versus $p_{T}$. Through projection and single Gaussian fit,
we also can get $\pi$ raw yields. Fig~\ref{fig3} (b) shows a fit
example in 1.4$<$$p_{T}$$<$1.6(GeV/$c$) where the pion
distribution start merging with those from koans and protons.

Counting the entries at the range
-0.1$<$$m^{2}$$<$0.1(GeV/$c^{2})^{2}$ was also used to compare
with the fitting yields. The difference between them was found
$<$$5\%$ in low $p_{T}$ range and was used as part of the
systematic uncertainty.

\section{Non-photonic and photonic background electrons}\label{composition}
The inclusive electron raw yields have three
components~\cite{Phenix_0609010}: (1)electrons from heavy-flavor
decay (charm and bottom), (2)photonic background electrons from
Dalitz decays of light mesons ($\pi$$^{0}$, $\eta$ etc.) and gamma
conversion. (3)other background electrons from $K_{e3}$ decays and
dielectron decays of vector mesons. Photonic background 2) is much
larger than other background, so we will use signed DCA (sDCA)
(distance of closest approach of a track in TPC to the interaction
point) to reject the electron background from gamma conversion at
high radius and cocktail method to remove background from Dalitz
decays of light mesons.


\begin{figure}[htbp]
\resizebox{0.5\textwidth}{!}{\includegraphics{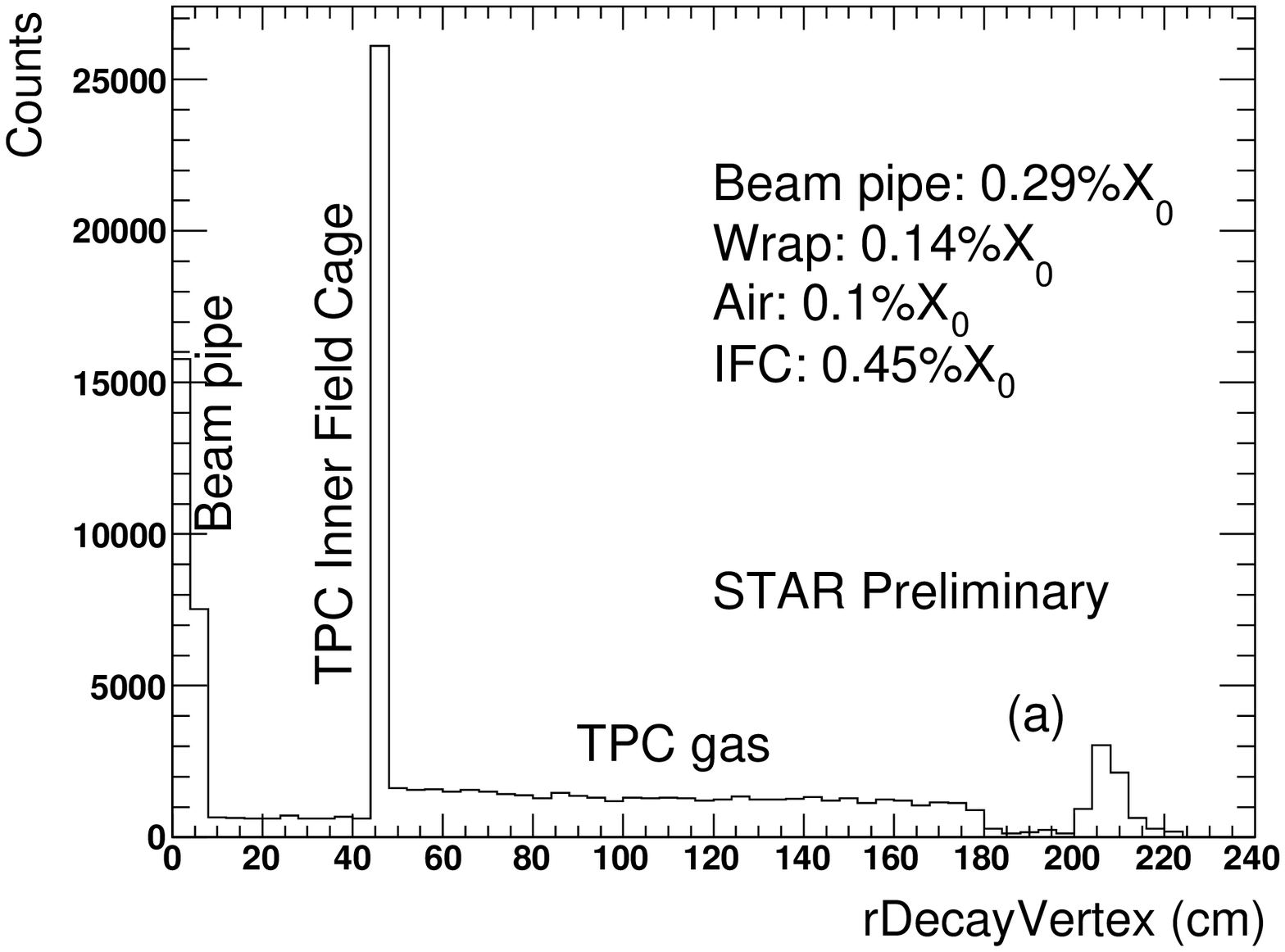}}
\resizebox{0.5\textwidth}{!}{\includegraphics{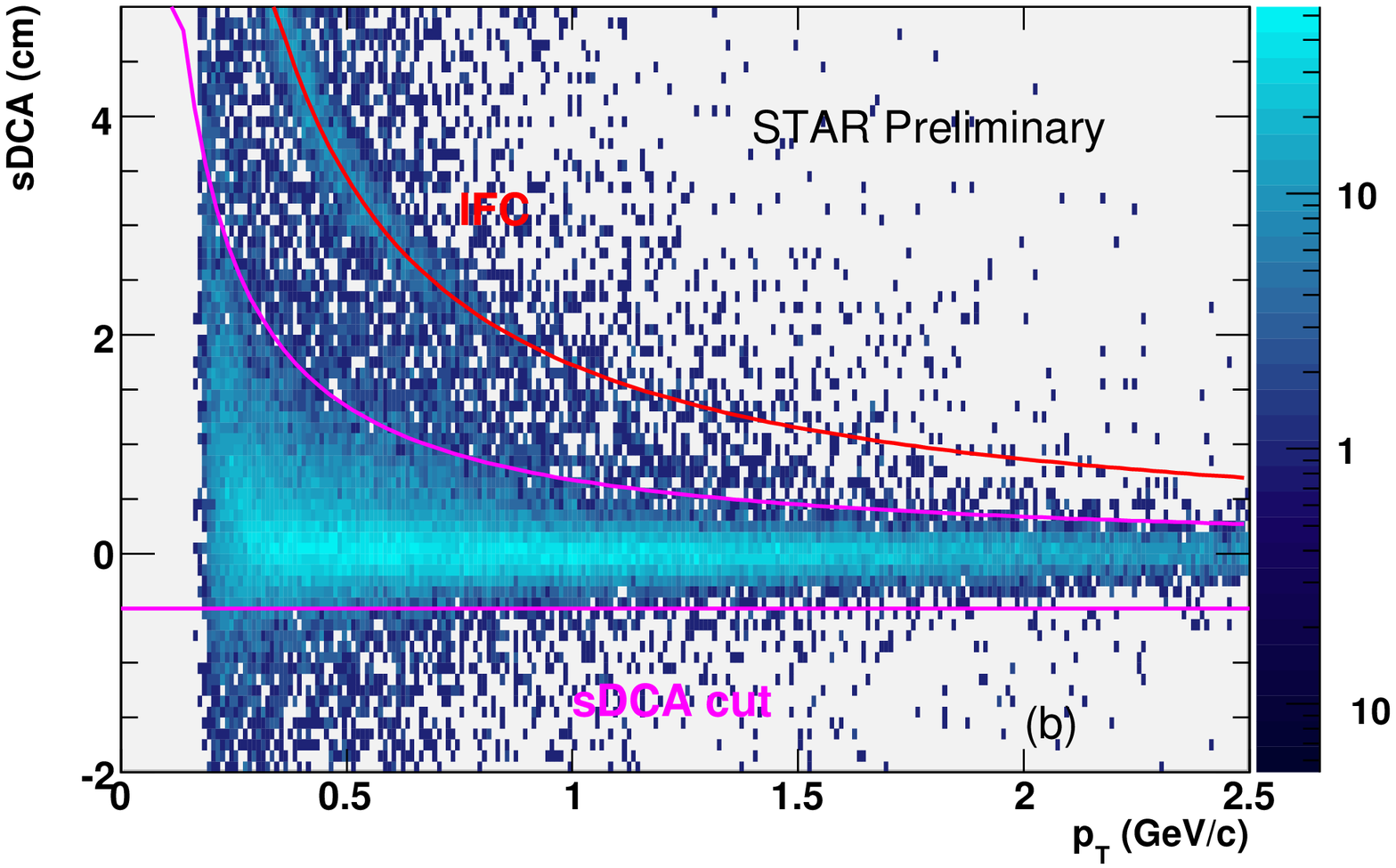}}
\caption[]{(Color online)(a) radial distance of gamma decay vertex to the primary vertex
from simulation.
(b) sDCA as a function of $p_{T}$. Lines are the sDCA for conversion at
TPC inner field cage (IFC) and the sDCA cuts.
} \label{fig4}
\end{figure}

Figure.~\ref{fig4} (a) shows radial distance (r) distribution of
gamma decay vertex to the primary vertex from a GEANT simulation.
There are two major background sources of gamma conversion,
material around the beam pipe ($Be$ beam pipe
$\scriptsize{\sim}$0.29\%$X_{0}$ + wraps for the beam pipe
bake-out) and TPC Inner Field Cage (IFC
$\scriptsize{\sim}$0.45\%$X_{0}$). Here, we used sDCA
 cut to
remove gamma conversion at high radius ($<$$30$cm).
Figure.~\ref{fig4} (b) shows the sDCA as a function of $p_{T}$
from run8 data. A hand calculation of where the sDCA should be
from conversions at the IFC agrees nicely with the band in the
data. sDCA = $\sqrt{(p_T/0.0015)^2+r^2}-(p_T/0.0015)$ where $r$ is
the $\gamma$ conversion radius in a uniform solenoidal magnetic
field of 0.5Tesla. We can use this expression to get the sDCA
value of sDCA1 when $r$=30cm. With -0.5$<$sDCA$<$sDCA1 cut, we
rejected electron from gamma conversion in the air with
$r$$>$30cm, at IFC and TPC gas.

After removing gamma conversion at high radius, we use cocktail
method to subtract background from Dalitz decays. A cocktail of
electron spectra from various background sources is calculated using
a Monto Carlo event Generator of hadron decays. The most important
background is the $\pi^{0}$ Dalitz decays. Through fit to the charge
$\pi$ spectra in non-singly diffractive (NSD) p+p collisions, we get
a function $B/(1+(m_{T}-m_{0})/nT)^{n}$ (in this expression, $m_{T}$
and $m_{0}$ stand for the particle transverse mass and rest mass
separately and it has three parameters: $B$, $n$ and $T$). With fixed
parameter $n$ = 9.7, this expression fit not only charge $\pi$ spectra
but also charge kaon, $K^{\star}$, $\rho$ and $\phi$ well, so we use
this expression as input to the generator. With cocktail method, we
get the electron background from Dalitz decays of light mesons.

\section{$e/\pi$ ratio}\label{ratio}

\begin{figure}[htbp]
\resizebox{0.5\textwidth}{!}{\includegraphics{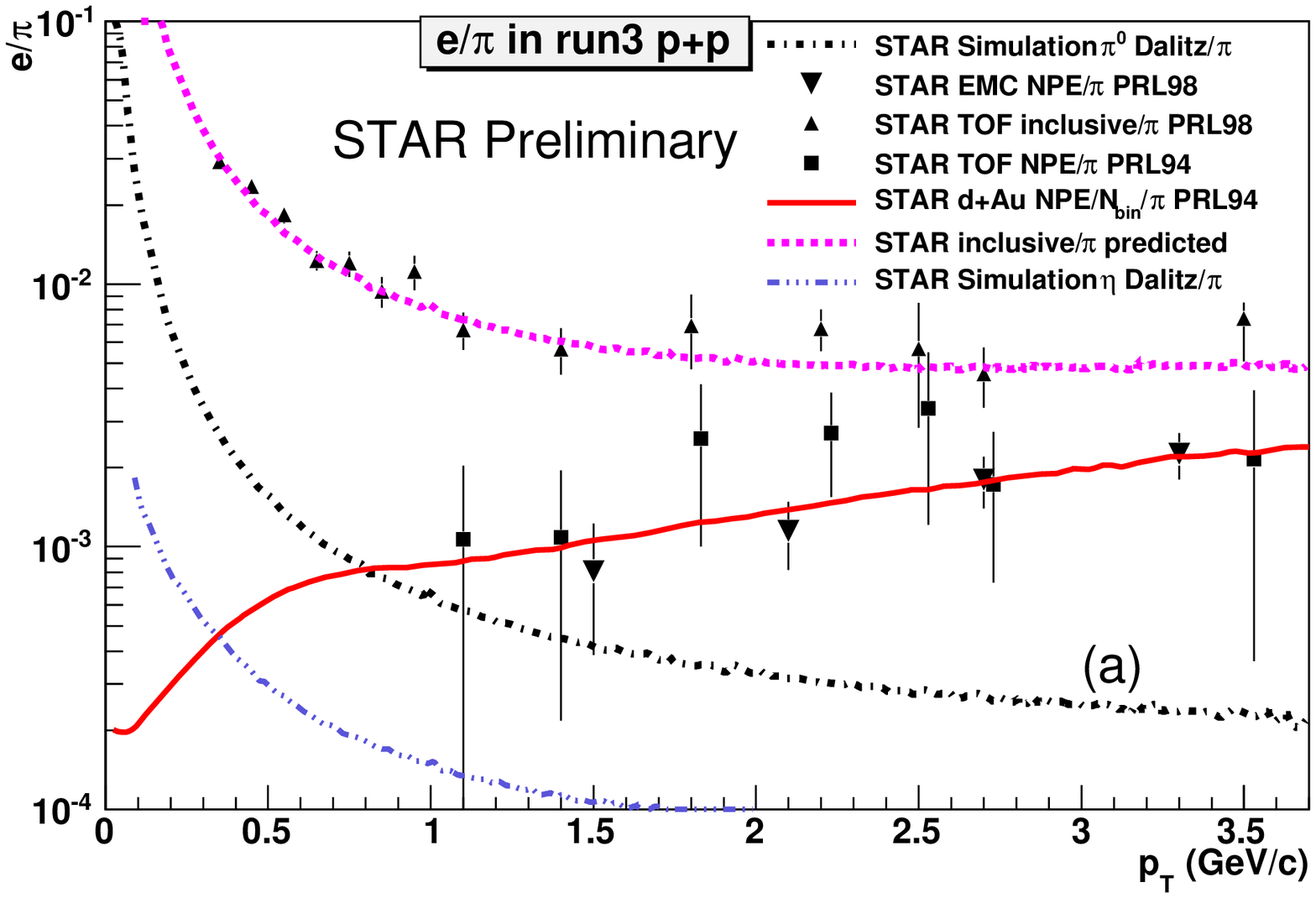}}
\resizebox{0.5\textwidth}{!}{\includegraphics{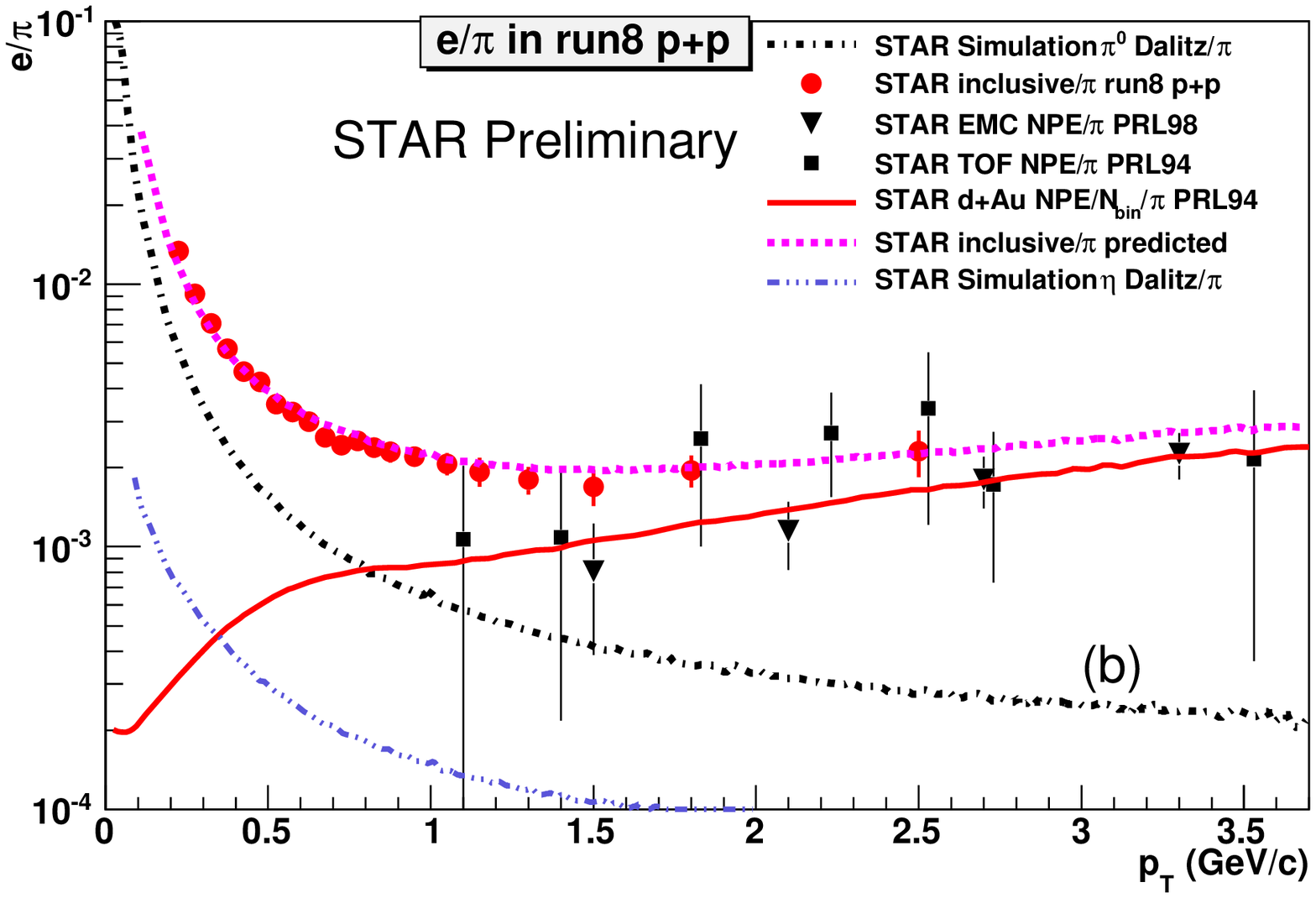}}
\caption[]{(Color online)(a) $e/\pi$ ratio as a function of $p_{T}$. The dashed line
is the sum of various background $e/\pi$ ratio ($\pi$$^{0}$/$\pi$, $\eta/\pi$
$\gamma$/$\pi$ and NPE/$\pi$) for run3 with an estimate of
material around the beam pipe to be a factor of x 10 higher than that in
run8. (b) Full circles represent the
inclusive $e/\pi$ ratio in run8 and the dashed line is similar as (a)
by requiring 0.69\%$X_{0}$ for gamma conversions.
} \label{fig5}
\end{figure}

Figure.~\ref{fig5} shows the $e/\pi$ ratio from run8 data, compared
to various background cocktails, NPE from previous results, and run3
inclusive electron to pion ratio. We also check the consistence of
the $e/\pi$ ratio from run8 and the results from run3. We take the
material budget from which $\gamma$ conversion in detector is
$\scriptsize{\sim}$10 in run3 than that in run8, and we include the
$e/\pi$ from $\pi^{0}$ and $e/\pi$ from $\eta$ Dalitz decays from
cocktail method and NPE/$\pi$ measured in run3 d+Au data scaled by
the binary collisions together, then we find the total sum of
$e/\pi$ from run8 is consistent with STAR TOF inclusive $e/\pi$ in
run3, and the $p_T$ dependence can be well reproduced as well. In
addition, $\gamma$ conversion is equivalent to 0.9 of the electron
yields from $\pi^{0}$ Dalitz decay in run8, comparable with the
estimated material budget in this run.

\section{Conclusions}
In summary, we present our analysis of mid-rapidity NPE production
at $p_{T}$$>$0.2GeV/$c$ in p+p collisions at $\sqrt{s_{NN}}$ = 200
GeV. Through the measurement of $e/\pi$ ratio, we find that the
photonic background electrons from gamma conversions are reduced
by about a factor of 10 compared with those in STAR previous runs
due to the absence of inner tracking detectors and the supporting
materials.
and preliminary results from run8
dataset agree with the results from run3.

\section{Acknowledgements}
This work was supported in part by the National Natural Science Foundation
of China under grant No. 10610285 and No.-10875159,
the Knowledge Innovation Project of the
Chinese Academy of Science under grant nos. KJCX2-YW-A14 and KJCX3-SYW-N2.

\end{document}